\documentclass[sigconf]{acmart}
\AtBeginDocument{%
  }

\copyrightyear{2026}
\acmYear{2026}
\setcopyright{cc}
\setcctype{by}
\acmConference[MM '26]{Proceedings of the 34th ACM International Conference on Multimedia}{November 10--14, 2026}{Rio de Janeiro, Brazil}
\acmBooktitle{Proceedings of the 34th ACM International Conference on Multimedia (MM '26), November 10--14, 2026, Rio de Janeiro, Brazil}
\acmDOI{10.1145/3767308.3836499}
\acmISBN{979-8-4007-2213-4/2026/11}

\begin{document}

\title[Orthogonal Purification and Topology-Guided MoE for Conflict-Aware Multimodal Recommendation]{Beyond Modality Harmony: Orthogonal Purification and Topology-Guided MoE for Conflict-Aware Multimodal Recommendation}

\author{Jialin Liu}
\email{camilla.liu@my.cityu.edu.hk}
\affiliation{%
  \institution{City University of Hong Kong}
  \city{Hong Kong}
  \country{Hong Kong}
}

\author{Zhaorui Zhang}
\authornote{Corresponding author.} 
\email{zhaorui.zhang@polyu.edu.hk}
\affiliation{%
  \institution{The Hong Kong Polytechnic University}
  \city{Hong Kong}
  \country{Hong Kong}
}

\author{Ray C. C. Cheung}
\email{r.cheung@cityu.edu.hk}
\affiliation{%
  \institution{City University of Hong Kong}
  \city{Hong Kong}
  \country{Hong Kong}
}
\renewcommand{\shortauthors}{Jialin Liu, Zhaorui Zhang, and Ray C. C. Cheung}

\begin{abstract}
Multimodal Recommender Systems (MRSs) typically rely on a flawed ``modality harmony'' assumption, presuming that multimodal features are inherently beneficial and strictly aligned with users' collaborative interaction patterns. However, modality-topology conflicts are ubiquitous in real-world scenarios due to deceptive visual clickbaits and mismatched semantics. Blindly integrating these noisy modalities inevitably pollutes the pristine collaborative space, causing severe representation distortion.  
To address this, we propose Orthogonal purification and topology-guided MoE for conflict-aware multimodal Recommendation (OrthoRec). At its core, OrthoRec introduces Collaborative-Guided Orthogonal Purification (CGOP), which geometrically decouples multimodal features into directions parallel and orthogonal to a pure collaborative anchor. By adaptively truncating the orthogonal noise with an energy-preserving normalization, CGOP rectifies deceptive semantic directions while preserving the modality's intrinsic representation capacity. Furthermore, we design a Topology-Aware Routing Mixture-of-Experts (TAR-MoE). Guided by the collaborative topology, TAR-MoE employs decoupled sigmoid gating to break the zero-sum bottleneck of traditional softmax attention, autonomously determining the injection scale for each purified modality. Finally, a safe-SSL objective is introduced to dynamically penalize the forced contrastive alignment of contradictory pairs. Experiments on three real-world Amazon datasets show that OrthoRec consistently outperforms competitive recent baselines and exhibits improved robustness under modality noise and item sparsity. Our code is available at \url{https://github.com/Camilla-jl/Orthorec}.
\end{abstract}





\maketitle

\section{Introduction}

\begin{figure}[t!]
\centering
\includegraphics[width=\columnwidth]{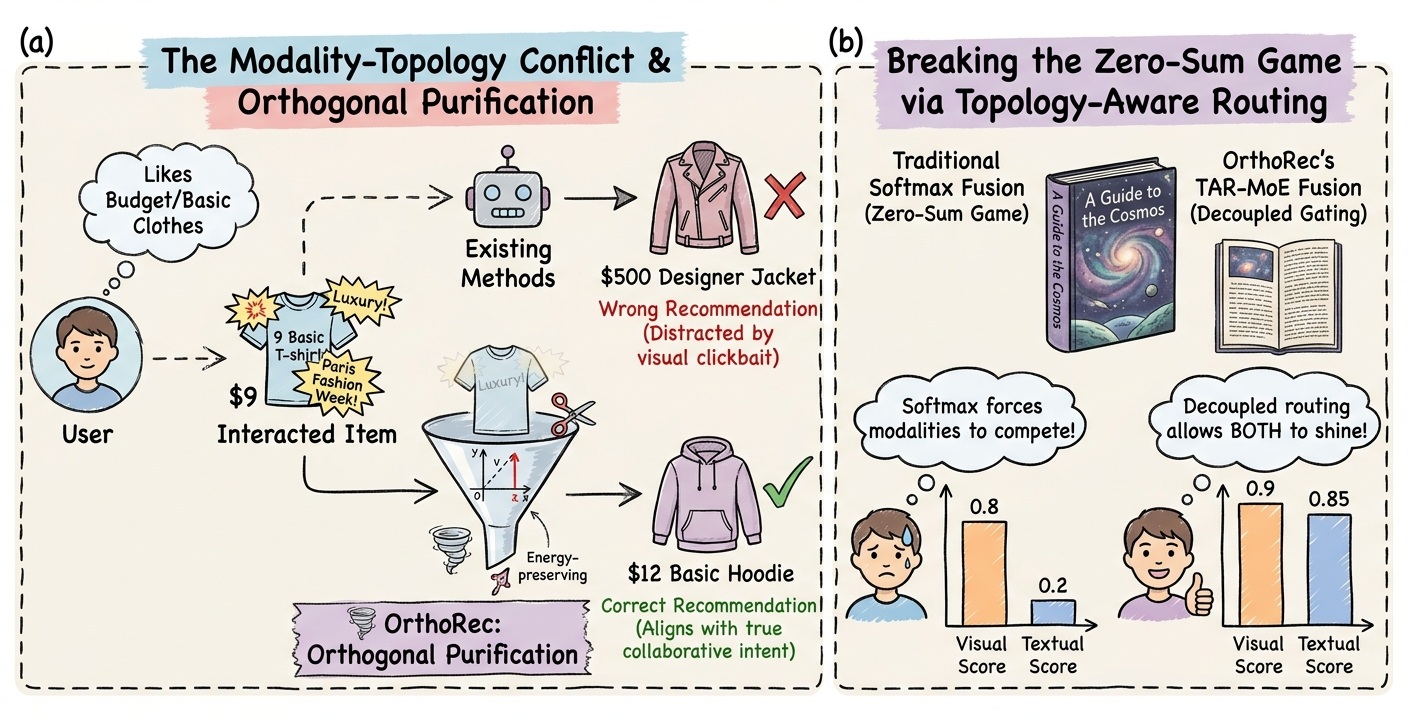}
\caption{Illustration of bottlenecks in current multimodal recommendation scenarios. (a) Impact of modality noise on user intent: blindly absorbing deceptive visual clickbaits distorts the true collaborative preference, leading to inaccurate recommendations. (b) Limitation of traditional modality fusion: softmax-based attention forces a zero-sum competition, heavily suppressing crucial textual cues when visual features are prominent.}
\label{fig:problem_illustration}
\end{figure}

Multimodal recommender systems (MRSs) have emerged as the foundational infrastructure for personalized information retrieval \cite{he2016vbpr, mcauley2015image, wei2019mmgcn, zhou2023bootstrap}. By incorporating visual and textual modalities, MRSs can effectively break the bottleneck of data sparsity and cold-start problems inherent in traditional collaborative filtering (CF) \cite{rendle2012bpr, zheng2021cold, liu2026sga}. Current state-of-the-art MRSs encompass a variety of approaches, such as graph-based message passing (e.g., LATTICE \cite{zhang2021mining}, MMGCN \cite{wei2019mmgcn}) and contrastive learning paradigms (e.g., MENTOR \cite{xu2025mentor}, BM3 \cite{zhou2023bootstrap}). The core philosophy behind these models is to integrate multimodal semantics with structural collaborative signals \cite{guo2024lgmrec, tao2022self}, aiming to construct a comprehensive and expressive latent representation for each user and item.

However, the prosperity of existing architectures implicitly relies on a fragile utopian assumption: \textbf{``Modality Harmony''} \cite{liu2024multimodal, zhou2023tale}. They presume that multimodal features are inherently flawless, universally beneficial, and strictly aligned with users' collaborative interaction patterns. In reality, this assumption is fundamentally flawed \cite{wang2024noise, zhang2022sedgn}. Real-world recommender systems are inundated with modality-topology conflicts, driven by visual clickbaits, exaggerated textual claims, or mismatched modality semantics \cite{jiang2024diffmm}. Such deceptive multimodal signals introduce false-positive edges into the user-item graph, fundamentally distorting the latent representation space \cite{gao2024causal, mu2022alleviating}. 

As illustrated in Figure \ref{fig:problem_illustration}, blindly absorbing these deceptive features triggers two critical bottlenecks in existing architectures. \textbf{First, destructive interference via blind alignment:} By forcefully aligning (via Contrastive Learning \cite{chen2020simple, yu2022graph}) or fusing (via GNNs \cite{he2020lightgcn, wei2020grcn}) these deceptive modalities with the topological graph, existing models inevitably pollute the pristine CF space \cite{zheng2021disentangling}. The user's true preference representation collapses into the spurious visual appeal, leading to severe latent space distortion \cite{wang2020understanding, robinson2020contrastive}. \textbf{Second, the zero-sum game in fusion:} Traditional fusion mechanisms typically rely on softmax-based attention \cite{vaswani2017attention} or heuristic summation \cite{wei2019mmgcn}. This enforces a zero-sum competition among modalities \cite{peng2022balanced} (e.g., increasing the visual weight strictly decreases the textual weight), completely ignoring the fact that modalities should be decoupled and independently evaluated based on the topological consensus of different items. To date, the question of how to systematically disentangle deceptive modality noise from true semantic intent \cite{zheng2021disentangling, tian2020makes} while preserving the modality's intrinsic representational capacity remains largely under-explored.

To bridge this gap, we propose \textbf{Ortho}gonal Purification and Topology-Guided MoE for Conflict-Aware Multimodal \textbf{Rec}ommendation (\textbf{OrthoRec}), a novel framework that transcends the modality harmony illusion via geometric decoupling. Our core insight is to treat the pure crowd-wisdom interaction patterns (i.e., the topological CF embeddings) as a trustworthy collaborative anchor. Rather than arbitrarily masking or discarding multimodal features, OrthoRec introduces a Collaborative-Guided Orthogonal Purification (CGOP) mechanism. We geometrically project the raw visual and textual features into directions parallel and orthogonal to the collaborative anchor. The parallel component retains the safe, consensus-aligned semantics, while the orthogonal component harboring both exploratory semantics and deceptive clickbait noise is adaptively scaled by a topology-aware conflict gate. Crucially, this purification is energy-preserving, since we dynamically restore the rectified vector's $L_2$ norm. This ensures that we only correct the deceptive ``semantic direction'' without artificially crippling the feature's inherent representation capacity.

Furthermore, to eliminate the zero-sum fusion bottleneck, we design a Topology-Aware Routing Mixture-of-Experts (TAR-MoE). Instead of using raw modalities to compute attention, TAR-MoE uses the topological collaborative anchor as the routing condition, employing decoupled sigmoid gating to independently determine the optimal injection scale for each purified modality. Coupled with a safe-SSL objective that dynamically down-weights the contrastive penalty for inherently contradictory modalities, OrthoRec effectively prevents latent space distortion.

The main contributions of this work are summarized as follows:
\begin{itemize}
    \item We explicitly challenge the ``modality harmony'' assumption by identifying the modality-topology conflict and the destructive interference caused by deceptive multimodal noise in MRSs.
    \item We propose an energy-preserving Orthogonal Purification (CGOP) mechanism. It provides a simple geometric approach to disentangle and rectify deceptive modality noise while alleviating magnitude decay.
    \item We introduce TAR-MoE with decoupled gating to break the zero-sum fusion bottleneck, and safe-SSL to mitigate the latent distortion caused by forced contrastive alignment.
    \item Experiments on three real-world Amazon datasets show that OrthoRec consistently outperforms competitive recent baselines, and exhibits improved robustness under injected modality noise and item sparsity.
\end{itemize}

\section{Related Work}

\subsection{Graph-based Multimodal Recommendation}
Multimodal Graph Neural Networks (GNNs) capture complex user-item connectivity and multimodal signals by constructing modality-specific graphs \cite{wei2019mmgcn, wei2020grcn, wang2021dualgnn, liu2026motorec} or mining latent semantic structures \cite{zhang2021mining, guo2024lgmrec}. However, these methods typically rely on early feature concatenation or heuristic summation, operating under a strict ``modality harmony'' assumption. This blind fusion makes them highly vulnerable to destructive interference from modality-topology conflicts (e.g., visual clickbaits). In contrast, OrthoRec abandons naive fusion by introducing a collaborative-guided orthogonal purification (CGOP) mechanism, geometrically filtering out deceptive modality noise prior to graph aggregation.

\subsection{Contrastive Alignment in Recommendation}
Contrastive learning (CL) is widely adopted to mitigate data sparsity by maximizing mutual information between multimodal views and ID embeddings \cite{tao2022self, zhou2023bootstrap, lin2025contrastive}, or by distilling modality-invariant representations via cross-modal alignment \cite{zhou2023tale, xu2025mentor}. While effective, existing CL methods blindly pursue semantic alignment across all items. When an item's visual and textual modalities are inherently contradictory, forcing alignment inevitably leads to latent space distortion, collapsing representations into spurious features \cite{wang2020understanding, robinson2020contrastive, yang2023debiased}. To address this, our proposed safe-SSL utilizes a geometric conflict score to dynamically down-weight the alignment penalty for contradictory pairs, ensuring a safe optimization space.

\subsection{Feature Denoising and Disentanglement}
To combat ubiquitous data noise, recent studies explore disentangled representation learning \cite{ma2019learning, zheng2021disentangling} and robust multimodal denoising \cite{guo2023noise, zhang2022sedgn, wang2024noise}. Nevertheless, existing techniques face two critical dilemmas: (1) heuristic masking or edge dropping \cite{rong2019dropedge} directly discards features, causing severe magnitude decay and representation capacity loss, and (2) traditional attention mechanisms \cite{vaswani2017attention} rely on zero-sum softmax gating, restricting the independent utilization of multiple modalities \cite{peng2022balanced}. OrthoRec addresses these bottlenecks by combining an energy-preserving orthogonal truncation (which rectifies deceptive semantic directions without shrinking the $L_2$ norm) with TAR-MoE \cite{ma2018modeling, shazeer2017outrageously}, which employs decoupled sigmoid gating to autonomously determine modality injection scales free from zero-sum constraints.

\begin{figure*}[t!] 
    \centering
    \includegraphics[width=0.9\textwidth]{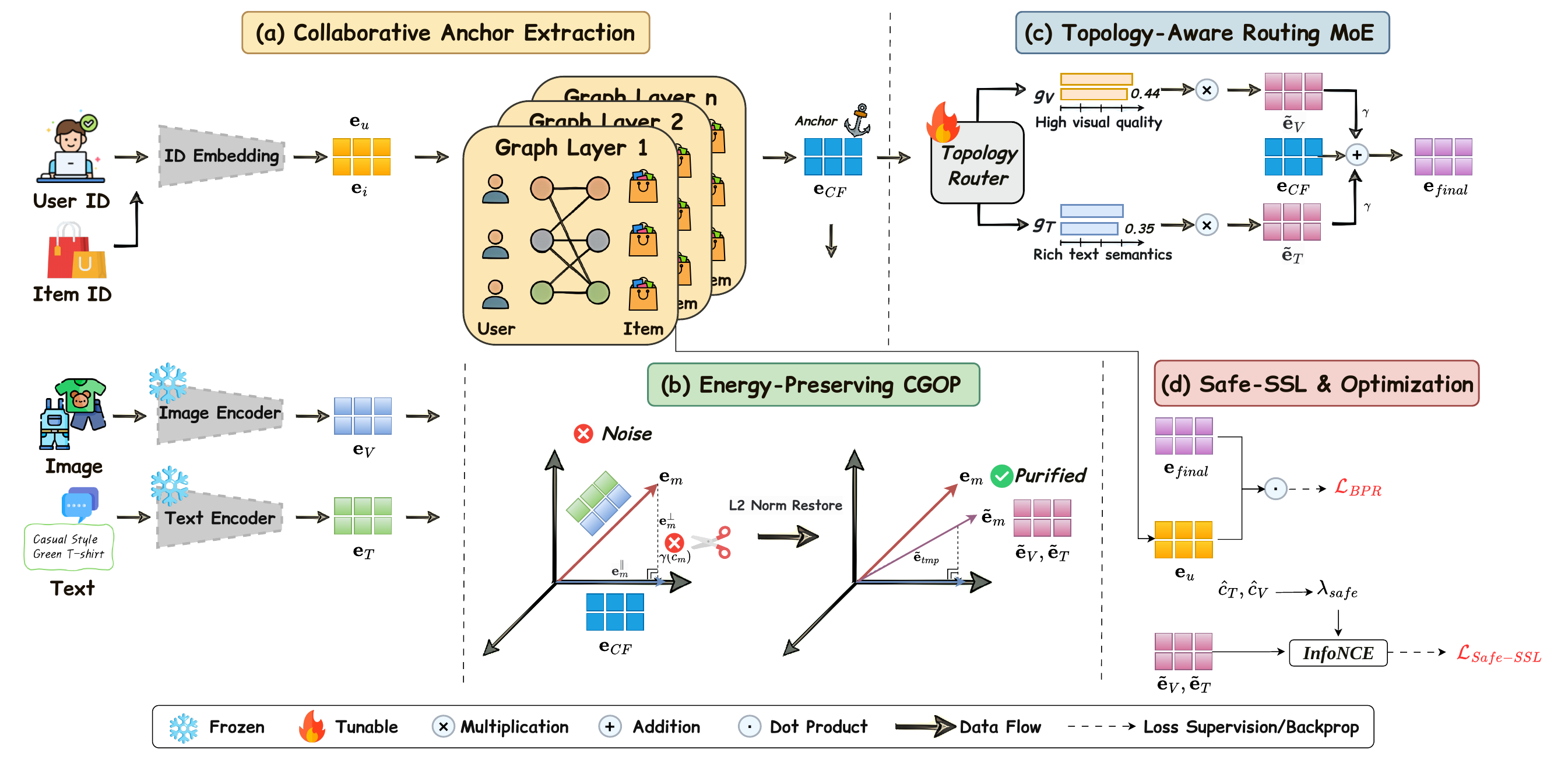}
    \caption{The overall architecture of OrthoRec. It consists of four main modules: (1) collaborative anchor extraction; (2) energy-preserving orthogonal purification to filter deceptive noise; (3) topology-aware routing MoE for decoupled fusion; and (4) safe-SSL and optimization, which prevents latent space distortion.}
    \label{fig:framework}
\end{figure*}

\section{Methodology}

\subsection{Problem Formulation}
Let $\mathcal{U}$ and $\mathcal{I}$ denote the sets of users and items, respectively. The historical user-item interactions are formally represented by a bipartite graph $\mathcal{G} = (\mathcal{U} \cup \mathcal{I}, \mathcal{E})$, where an edge $(u, i) \in \mathcal{E}$ indicates that user $u$ has interacted with item $i$. For multimodal side information, each item $i \in \mathcal{I}$ is associated with raw visual and textual features extracted from pre-trained encoders, denoted as $\mathbf{x}_V^{(i)} \in \mathbb{R}^{d_V}$ and $\mathbf{x}_T^{(i)} \in \mathbb{R}^{d_T}$.
To project these heterogeneous modalities into a unified latent space, we employ modality-specific linear transformations:
\begin{equation}
    \mathbf{e}_m^{(i)} = \mathbf{W}_m \mathbf{x}_m^{(i)} + \mathbf{b}_m, \quad m \in \{V, T\}
\end{equation}
where $\mathbf{W}_m \in \mathbb{R}^{d \times d_m}$ and $\mathbf{b}_m \in \mathbb{R}^d$ are trainable weight matrices and biases, transforming the raw features into $d$-dimensional dense embeddings $\mathbf{e}_V^{(i)}$ and $\mathbf{e}_T^{(i)}$.
Given the interaction graph $\mathcal{G}$ and the unified multimodal features $\{\mathbf{e}_V^{(i)}, \mathbf{e}_T^{(i)}\}$, our ultimate goal is to learn robust, noise-resistant representations for users and items. These purified representations are then utilized to accurately predict the unobserved interaction probability $\hat{y}_{u,i}$.

\subsection{Collaborative Anchor Extraction}
\label{sec:anchor}
Most existing multimodal recommenders adopt an early-fusion strategy, blindly injecting visual and textual features into the graph message passing at the very beginning \cite{wei2019mmgcn, zhang2021mining, guo2024lgmrec}. However, we argue that this conventional practice inevitably corrupts the topological structure with modality noise (e.g., visual clickbaits or mismatched semantics) \cite{wang2024noise}. Before aligning or fusing any heterogeneous modalities, it is imperative for the model to first establish a pristine, trustworthy structural reference.

To achieve this, we exclusively perform graph convolutions on the pure ID embeddings over $\mathcal{G}$, isolating them from any multimodal interference. We initialize trainable ID embeddings $\mathbf{e}_u^{(0)} \in \mathbb{R}^d$ and $\mathbf{e}_i^{(0)} \in \mathbb{R}^d$ for users and items. Following the standard LightGCN architecture \cite{he2020lightgcn}, the message passing paradigm at the $l$-th layer is defined as:
\begin{equation}
    \mathbf{e}_u^{(l)} = \sum_{i \in \mathcal{N}_u} \frac{1}{\sqrt{|\mathcal{N}_u||\mathcal{N}_i|}} \mathbf{e}_i^{(l-1)}, \quad
    \mathbf{e}_i^{(l)} = \sum_{u \in \mathcal{N}_u} \frac{1}{\sqrt{|\mathcal{N}_i||\mathcal{N}_u|}} \mathbf{e}_u^{(l-1)}
\end{equation}
where $\mathcal{N}_u$ and $\mathcal{N}_i$ denote the first-order topological neighbors of user $u$ and item $i$, respectively. The symmetric normalization term $1/\sqrt{|\mathcal{N}_u||\mathcal{N}_i|}$ serves to discount the impact of high-degree nodes, preventing representation over-smoothing. After $L$ layers of propagation, we aggregate the embeddings across all layers to obtain the final ID-based representation:
\begin{equation}
    \mathbf{e}_i^{ID} = \frac{1}{L+1} \sum_{l=0}^{L} \mathbf{e}_i^{(l)}
\end{equation}

We define this resulting representation $\mathbf{e}_i^{ID}$ as the collaborative anchor ($\mathbf{e}_{CF} \in \mathbb{R}^d$). Following LGMRec \cite{guo2024lgmrec}, we additionally enhance the anchor with a global hypergraph embedding: $\mathbf{e}_{CF} \leftarrow \mathbf{e}_{CF} + \alpha \cdot \mathbf{e}^{GHE}$, where $\mathbf{e}^{GHE}$ is the normalized hypergraph-propagated embedding and $\alpha$ is the hypergraph enhancement weight. This design inherits the ability to capture global structural dependencies, while our contributions operate on top of this anchor. By distilling the pure ``crowd-wisdom'' interaction patterns, $\mathbf{e}_{CF}$ serves as a robust collaborative consensus. It provides the essential geometric reference axis required to detect and rectify deceptive modality noise in the subsequent orthogonal purification stage.

\subsection{Collaborative-Guided Orthogonal Purification}
\label{sec:cgop}

Given the reliable collaborative anchor $\mathbf{e}_{CF}$, we discard the naive assumption that raw multimodal features $\mathbf{e}_m$ ($m \in \{T, V\}$) are flawless. In real-world scenarios plagued by visual clickbaits or exaggerated textual claims, $\mathbf{e}_m$ often encapsulates deceptive noise that strictly conflicts with the user's true behavioral intent \cite{wang2024noise}. To address this, we propose Collaborative-Guided Orthogonal Purification (CGOP), a module that geometrically decouples $\mathbf{e}_m$ into a safe consensus direction and an exploratory/noisy direction.

\textbf{Geometric Decoupling.} Taking the collaborative embedding $\mathbf{e}_{CF}$ as the structural reference axis, we project the raw multimodal feature $\mathbf{e}_m$ onto it to extract the consistent parallel component:
\begin{equation}
    \mathbf{e}_{m}^{\parallel} = \left( \frac{\mathbf{e}_m \cdot \mathbf{e}_{CF}}{||\mathbf{e}_{CF}||_2^2} \right) \mathbf{e}_{CF}
\end{equation}
where $(\cdot)$ denotes the inner product and $||\cdot||_2$ is the $L_2$ norm. This parallel component precisely captures the modality semantics that strictly align with the topological behaviors verified by the crowd. Consequently, the residual forms the orthogonal component:
\begin{equation}
    \mathbf{e}_{m}^{\perp} = \mathbf{e}_m - \mathbf{e}_{m}^{\parallel}
\end{equation}
By definition, the orthogonal component $\mathbf{e}_{m}^{\perp}$ lies in a null space mathematically independent of the collaborative graph. It harbors modality-specific unique information, encompassing both beneficial long-tail exploratory semantics and deceptive clickbait noise.

\textbf{Conflict-Aware Adaptive Truncation.} To selectively filter out the deceptive noise within the orthogonal space, we compute a cosine-based conflict score $c_m \in [-1, 1]$:
\begin{equation}
    c_m = \frac{\mathbf{e}_m \cdot \mathbf{e}_{CF}}{||\mathbf{e}_m||_2 ||\mathbf{e}_{CF}||_2}
\end{equation}
A lower (or negative) $c_m$ indicates a severe modality-topology conflict. In practice, we center $c_m$ by subtracting its batch mean $\bar{c}_m$ before gating, so that the gate responds to an item's \textit{relative} conflict severity rather than the global modality gap between pre-trained features and the CF space. Crucially, this conflict score acts as a universal indicator of semantic trustworthiness. It not only governs the local feature truncation within CGOP but also serves as the fundamental guidance for the global safe alignment (detailed in Section \ref{sec:safessl}). We dynamically scale the orthogonal space via a soft gating function $\gamma(c_m) = \sigma(g_\theta(c_m))$, where $g_\theta(\cdot)$ is a modality-shared lightweight two-layer MLP and $\sigma(\cdot)$ is the sigmoid activation. The temporarily truncated feature is thus formulated as:
\begin{equation}
    \tilde{\mathbf{e}}_{tmp} = \mathbf{e}_{m}^{\parallel} + \gamma(c_m) \cdot \mathbf{e}_{m}^{\perp}
\end{equation}

\textbf{Energy-Preserving Normalization.} In high-dimensional representation learning, the magnitude (i.e., $L_2$ norm) of an embedding vector typically correlates with its feature expressiveness and confidence \cite{wang2020understanding}. Directly outputting the truncated feature $\tilde{\mathbf{e}}_{tmp}$ would lead to severe magnitude decay, artificially shrinking the modality's intrinsic representation capacity. To preserve the representation magnitude, we introduce an energy-preserving normalization:
\begin{equation}
    \tilde{\mathbf{e}}_m = \frac{\tilde{\mathbf{e}}_{tmp}}{||\tilde{\mathbf{e}}_{tmp}||_2} \cdot ||\mathbf{e}_m||_2
\end{equation}
Through this geometric operation, CGOP rectifies the noisy modality's deceptive semantic direction while retaining its representational energy. In essence, it reorients the modality vector away from topological conflicts without artificially shrinking its magnitude. Having obtained the purified multimodal features $\tilde{\mathbf{e}}_V$ and $\tilde{\mathbf{e}}_T$, the next critical step is to dynamically aggregate them with the collaborative anchor $\mathbf{e}_{CF}$ to form a unified item representation.

\subsection{Topology-Aware Routing MoE (TAR-MoE)}
\label{sec:tarmoe}
To perform the aforementioned aggregation, traditional methods typically rely on heuristic summation or softmax-based attention mechanisms \cite{vaswani2017attention}. However, softmax inherently traps modalities in a zero-sum game \cite{peng2022balanced}, assigning a higher weight to the visual feature strictly suppresses the textual counterpart. This rigid competition contradicts real-world recommendation scenarios where an item might offer equally crucial visual and textual cues, demanding high injection scales for both simultaneously.

To break this zero-sum bottleneck, we propose TAR-MoE equipped with decoupled sigmoid gating \cite{ma2018modeling}. We conceptualize the fusion process as a non-mutually-exclusive routing mechanism, where the local and global graph topology ($\mathbf{e}_{CF}$) serves as the ultimate condition to independently evaluate each modality. The routing weight $g_m \in (0, 1)$ for each modality is calculated as:
\begin{equation}
    g_m = \sigma\!\left(\mathrm{MLP}_{route\_m}(\mathbf{e}_{CF}) / \tau_r\right), \quad m \in \{T, V\}
\end{equation}
where $\tau_r$ is a routing temperature and $\sigma(\cdot)$ is the sigmoid function. The final unified item representation $\mathbf{e}_{final}$ is then dynamically aggregated via:
\begin{equation}
    \mathbf{e}_{final} = \mathbf{e}_{CF} + \gamma \left( g_{T} \cdot \tilde{\mathbf{e}}_{T} + g_{V} \cdot \tilde{\mathbf{e}}_{V} \right)
\end{equation}
where $\gamma$ is a global modality injection scale balancing the purified multimodal signals against the collaborative anchor (note that $\gamma$ is distinct from the item-wise truncation gate $\gamma(c_m)$ in CGOP). In practice, the purified features are passed through identity-initialized expert layers and $L_2$-normalized before injection, which stabilizes the injected signal magnitudes. Unlike previous attention mechanisms computed from the raw, potentially noisy modalities themselves, our decoupled routing strategy empowers the pure graph structure to autonomously determine the injection scale for each purified signal. This design allows the model to fully capitalize on all beneficial multimodal features without destructive interference.

\subsection{Conflict-Aware Safe Contrastive Learning}
\label{sec:safessl}
While TAR-MoE effectively fuses multimodal features for the primary recommendation task, it is also crucial to align the underlying semantic spaces of different modalities. To this end, cross-modal contrastive learning (SSL) is widely adopted as an auxiliary task to maximize the mutual information between paired texts and images \cite{tao2022self}.

\textbf{The Latent Distortion Dilemma.} Traditional SSL blindly pursues semantic alignment across all items. However, this introduces a critical vulnerability: \textit{False Positive Alignment}. When an item's visual and textual modalities are inherently contradictory (e.g., a factual text paired with a deceptive clickbait image), forcing them to align with each other inevitably leads to latent space distortion \cite{wang2020understanding}.

\textbf{Safe-SSL Objective.} To ensure a safe optimization boundary, we propose to dynamically penalize the contrastive alignment for inherently contradictory modality pairs. Rather than introducing external heuristics, we directly reuse CGOP's own conflict estimates inside the contrastive objective. We seamlessly reuse the centered geometric conflict scores $\hat{c}_T, \hat{c}_V$ (i.e., $c_m$ after subtracting its batch mean, as detected during the CGOP stage in Section \ref{sec:cgop}). We design a dynamic penalty weight $\lambda_{safe}^{(i)}$ for each item $i$:
\begin{equation}
    \lambda_{safe}^{(i)} = \sigma\left( \frac{\hat{c}_T^{(i)} + \hat{c}_V^{(i)}}{\tau_{safe}} \right)
\end{equation}
where $\sigma(\cdot)$ is the sigmoid function and $\tau_{safe}$ is a temperature hyperparameter controlling the penalty sharpness.

Since $\sigma(\cdot)$ is bounded, $\lambda_{safe}^{(i)}$ lies within $(0, 1)$. If both modalities are relatively harmonious with the collaborative anchor ($\hat{c}_T, \hat{c}_V > 0$), $\lambda_{safe}$ approaches $1$ (full alignment). Conversely, if either modality conflicts with the topology substantially more than average (e.g., $\hat{c}_V \ll 0$ due to visual noise), $\lambda_{safe}$ decays smoothly towards $0$.

The standard InfoNCE loss is then reformulated into our safe-SSL loss:
\begin{equation}
    \mathcal{L}_{Safe-SSL} = - \sum_{i \in \mathcal{B}} \lambda_{safe}^{(i)} \ln \frac{\exp(\text{sim}(\tilde{\mathbf{e}}_{T}^{(i)}, \tilde{\mathbf{e}}_{V}^{(i)})/\tau_{ssl})}{\sum_{j \in \mathcal{B}} \exp(\text{sim}(\tilde{\mathbf{e}}_{T}^{(i)}, \tilde{\mathbf{e}}_{V}^{(j)})/\tau_{ssl})}
\end{equation}
where $\mathcal{B}$ denotes the mini-batch, $\text{sim}(\cdot, \cdot)$ is the cosine similarity function, and $\tau_{ssl}$ is the contrastive temperature. The denominator sums over all in-batch negative samples $j$.

By integrating this dynamic penalty, safe-SSL acts as an automatic safety valve. It instructs the model to confidently reject the alignment of toxic pairs, thereby protecting the purified semantic space from optimization distortion.

\subsection{Model Optimization}
The overall framework is optimized end-to-end in a multi-task learning manner. For the primary recommendation task, we employ the widely-used Bayesian Personalized Ranking (BPR) loss, which encourages the model to rank observed positive items higher than unobserved negative ones:
\begin{equation}
    \mathcal{L}_{BPR} = - \sum_{(u,i,j) \in \mathcal{O}} \ln \sigma(\hat{y}_{u,i} - \hat{y}_{u,j})
\end{equation}
where $\mathcal{O} = \{(u, i, j) \mid i \in \mathcal{I}_u^+, j \in \mathcal{I} \setminus \mathcal{I}_u^+\}$ denotes the set of pairwise training triplets, with $\mathcal{I}_u^+$ representing the items interacted by user $u$, and $j$ sampled as an unobserved negative item. In practice we additionally apply a small negative margin $m$ inside the sigmoid (i.e., $\sigma(\hat{y}_{u,i} - \hat{y}_{u,j} - m)$) to encourage a larger separation, where $m$ is grid-searched on the validation set. The predicted preference score is calculated via the inner product $\hat{y}_{u,i} = \mathbf{e}_u^\top \mathbf{e}_{final, i}$, where $\mathbf{e}_u$ is the final user representation obtained from the graph encoder.

To jointly optimize the primary recommendation task and the conflict-aware semantic alignment, the overall objective function is formulated as:
\begin{equation}
    \mathcal{L}_{total} = \mathcal{L}_{BPR} + \lambda_{cl} \mathcal{L}_{Safe-SSL} + \lambda_{reg} \|\Theta\|_2^2
\end{equation}
where $\lambda_{cl}$ is a hyperparameter controlling the safe contrastive learning strength (the safe-SSL weight, reported as the CL weight in Section~\ref{sec:rq5}), $\lambda_{reg}$ is the weight for $L_2$ regularization to prevent overfitting, and $\Theta$ denotes all trainable parameters in the model.

\section{Experiments}

We conduct extensive experiments on three real-world datasets to comprehensively evaluate the proposed OrthoRec framework. 

\subsection{Experimental Setup}
\begin{table}[t!] 
\centering
\caption{Statistics of the experimental datasets.}
\renewcommand{\arraystretch}{0.9}
\begin{tabular}{lrrrr} 
\hline
\hline
\textbf{Dataset} & \textbf{\#Users} & \textbf{\#Items} & \textbf{\#Inters.} & \textbf{Sparsity} \\
\hline
Baby & 19,445 & 7,050 & 160,792 & 99.88\% \\
Sports & 35,598 & 18,357 & 296,337 & 99.95\% \\
Clothing & 39,387 & 23,033 & 278,677 & 99.97\% \\
\hline
\hline
\end{tabular}
\label{tab:datasets}
\end{table}

\begin{table*}[t]
\centering
\caption{Overall performance comparison on three datasets. The best results are highlighted in \textbf{bold}, and the strongest baseline results are \underline{underlined}. ``Improv.'' denotes OrthoRec's relative improvement over the best baseline.}
\label{tab:overall_performance}
\renewcommand{\arraystretch}{0.9}
\resizebox{\textwidth}{!}{
\begin{tabular}{l|cccc|cccc|cccc}
\toprule\toprule
\textbf{Datasets} & \multicolumn{4}{c|}{\textbf{Baby}} & \multicolumn{4}{c|}{\textbf{Sports}} & \multicolumn{4}{c}{\textbf{Clothing}} \\
\cline{2-13}
\textbf{Models} & R@10 & R@20 & N@10 & N@20 & R@10 & R@20 & N@10 & N@20 & R@10 & R@20 & N@10 & N@20 \\
\midrule
MF-BPR  & 0.0357 & 0.0575 & 0.0192 & 0.0249 & 0.0432 & 0.0653 & 0.0241 & 0.0298 & 0.0187 & 0.0279 & 0.0103 & 0.0126 \\
LightGCN & 0.0479 & 0.0754 & 0.0257 & 0.0328 & 0.0569 & 0.0864 & 0.0311 & 0.0387 & 0.0340 & 0.0526 & 0.0188 & 0.0236 \\
SimGCL  & 0.0513 & 0.0804 & 0.0273 & 0.0350 & 0.0601 & 0.0919 & 0.0327 & 0.0414 & 0.0356 & 0.0549 & 0.0195 & 0.0244 \\
LayerGCN & 0.0529 & 0.0820 & 0.0281 & 0.0355 & 0.0594 & 0.0916 & 0.0323 & 0.0406 & 0.0371 & 0.0566 & 0.0200 & 0.0247 \\
\midrule
VBPR    & 0.0423 & 0.0663 & 0.0223 & 0.0284 & 0.0558 & 0.0856 & 0.0307 & 0.0384 & 0.0281 & 0.0415 & 0.0158 & 0.0192 \\
MMGCN   & 0.0378 & 0.0615 & 0.0200 & 0.0261 & 0.0370 & 0.0605 & 0.0193 & 0.0254 & 0.0218 & 0.0345 & 0.0110 & 0.0142 \\
DualGNN & 0.0448 & 0.0716 & 0.0240 & 0.0309 & 0.0568 & 0.0859 & 0.0310 & 0.0385 & 0.0454 & 0.0683 & 0.0241 & 0.0299 \\
SLMRec  & 0.0529 & 0.0775 & 0.0290 & 0.0353 & 0.0663 & 0.0990 & 0.0365 & 0.0450 & 0.0452 & 0.0675 & 0.0247 & 0.0303 \\
GRCN    & 0.0536 & 0.0829 & 0.0286 & 0.0363 & 0.0609 & 0.0925 & 0.0323 & 0.0408 & 0.0431 & 0.0661 & 0.0229 & 0.0278 \\
LATTICE & 0.0547 & 0.0850 & 0.0292 & 0.0370 & 0.0620 & 0.0953 & 0.0335 & 0.0421 & 0.0492 & 0.0733 & 0.0268 & 0.0330 \\
FREEDOM & 0.0627 & \underline{0.0992} & 0.0330 & 0.0424 & 0.0717 & \underline{0.1089} & 0.0385 & \underline{0.0481} & \underline{0.0628} & \underline{0.0941} & \underline{0.0341} & \underline{0.0420} \\
BM3     & 0.0564 & 0.0883 & 0.0301 & 0.0383 & 0.0656 & 0.0980 & 0.0355 & 0.0438 & 0.0422 & 0.0621 & 0.0231 & 0.0281 \\
MMGCL     & 0.0522 & 0.0779 & 0.0288 & 0.0357 & 0.0660 & 0.0992 & 0.0359 & 0.0445 & 0.0433 & 0.0667 & 0.0239 & 0.0291 \\
LGMRec  & \underline{0.0639} & 0.0989 & \underline{0.0337} & \underline{0.0430} & \underline{0.0719} & 0.1068 & \underline{0.0387} & 0.0477 & 0.0555 & 0.0828 & 0.0302 & 0.0371 \\
DA-MRS  & 0.0561 & 0.0895 & 0.0302 & 0.0388 & 0.0608 & 0.0952 & 0.0322 & 0.0410 & 0.0518 & 0.0779 & 0.0278 & 0.0344 \\
\midrule
\textbf{OrthoRec} & \textbf{0.0695} & \textbf{0.1057} & \textbf{0.0375} & \textbf{0.0466} & \textbf{0.0762} & \textbf{0.1158} & \textbf{0.0421} & \textbf{0.0514} & \textbf{0.0658} & \textbf{0.0955} & \textbf{0.0352} & \textbf{0.0436} \\
\midrule
Improv. & +8.76\% & +6.55\% & +11.28\% & +8.37\% & +5.98\% & +6.34\% & +8.79\% & +6.86\% & +4.78\% & +1.49\% & +3.23\% & +3.81\% \\
\bottomrule\bottomrule
\end{tabular}
}
\end{table*}

\textbf{Datasets.} We evaluate our model on three widely-used public datasets from the Amazon Product Reviews \cite{mcauley2015image}: \textit{Baby}, \textit{Sports and Outdoors}, and \textit{Clothing, Shoes and Jewelry}. Detailed statistics are presented in Table~\ref{tab:datasets}. To ensure a fair comparison, we follow the same data processing and filtering settings as in prior works that use these common benchmark datasets \cite{zhou2023tale, zhou2023bootstrap}. To represent the visual and textual modalities, we utilize pre-trained BEiT \cite{wang2023image} and BGE \cite{chen2024bge} sentence embeddings, respectively, across all datasets. 

\textbf{Evaluation Metrics.} We use an 8:1:1 train/validation/test split for all interactions. To evaluate the ranking performance, we use two standard metrics: Recall@$K$ and Normalized Discounted Cumulative Gain (NDCG@$K$), where $K \in \{10, 20\}$. We report the average metrics across all test users.

\textbf{Baselines.} We compare OrthoRec with three groups of representative baselines:
    \textit{General CF Models:} \textbf{MF-BPR} \cite{rendle2012bpr}, \textbf{LightGCN} \cite{he2020lightgcn}, \textbf{SimGCL} \cite{yu2022graph} and \textbf{LayerGCN} \cite{zhou2023layer}, which solely rely on user-item interaction structures.
    \textit{Graph-based Multimodal Models:} \textbf{VBPR} \cite{he2016vbpr}, \textbf{MMGCN} \cite{wei2019mmgcn}, \textbf{GRCN} \cite{wei2020grcn}, \textbf{DualGNN} \cite{wang2021dualgnn}, \textbf{LATTICE} \cite{zhang2021mining} and \textbf{LGMRec} \cite{guo2024lgmrec}. These models incorporate multimodal features into graph propagation using early fusion, modality-aware graphs, or structural refinement.
    \textit{Contrastive Learning \& Denoising Models:} \textbf{SLMRec} \cite{tao2022self}, \textbf{BM3} \cite{zhou2023bootstrap}, \textbf{FREEDOM} \cite{zhou2023tale}, \textbf{MMGCL} \cite{jiang2024mmgcl} and \textbf{DA-MRS} \cite{xv2024damrs}. These SOTA models use cross-modal contrastive learning or denoising-and-alignment objectives to purify multimodal semantics and user feedback.

\textbf{Implementation Details.}
OrthoRec is implemented in PyTorch. The latent dimension $d$ is set to $64$ on \textit{Baby} and to $128$ on \textit{Sports} and \textit{Clothing}. We optimize via Adam (batch size $4096$) with early stopping based on validation Recall@20, using a patience of $150$ epochs. Key hyperparameters (e.g., learning rate, $L_2$ weight, $\lambda_{cl}$) are tuned via grid search on the validation set, and the main hyperparameters' sensitivity is reported in Section~\ref{sec:rq5}. 


\subsection{Overall Performance (RQ1)}
Table \ref{tab:overall_performance} summarizes the comparison. \textbf{(1) OrthoRec improves consistently over all baselines}, ranking first on every metric of every dataset, with gains up to 11.28\%. At $K{=}20$ the NDCG gain exceeds the Recall gain throughout, so purification not only retrieves more relevant items but ranks them higher, as expected when deceptive directions are removed. \textbf{(2) Blind fusion can be actively harmful.} Several multimodal baselines fall \textit{below} the strongest pure-CF model despite access to strictly more information. Multimodal signal is thus not free, and the ``Modality Harmony'' assumption is violated often enough to cost these models the benefit of their features, the premise CGOP targets by truncating deceptive directions without shrinking representational capacity. \textbf{(3) The strongest baselines hedge against noise.} FREEDOM and LGMRec lead the field, both through structural rather than semantic treatment: a frozen denoised item-item graph and a global hypergraph. Their edge over unguarded early fusion confirms that noise must be handled, while their gap to OrthoRec shows structural hedging is no substitute for filtering the features themselves and routing them without zero-sum competition, as corroborated in Section \ref{sec:ablation}.

\subsection{Ablation Study (RQ2)}
\label{sec:ablation}

\begin{figure}[t!] 
    \centering
    \includegraphics[width=\columnwidth]{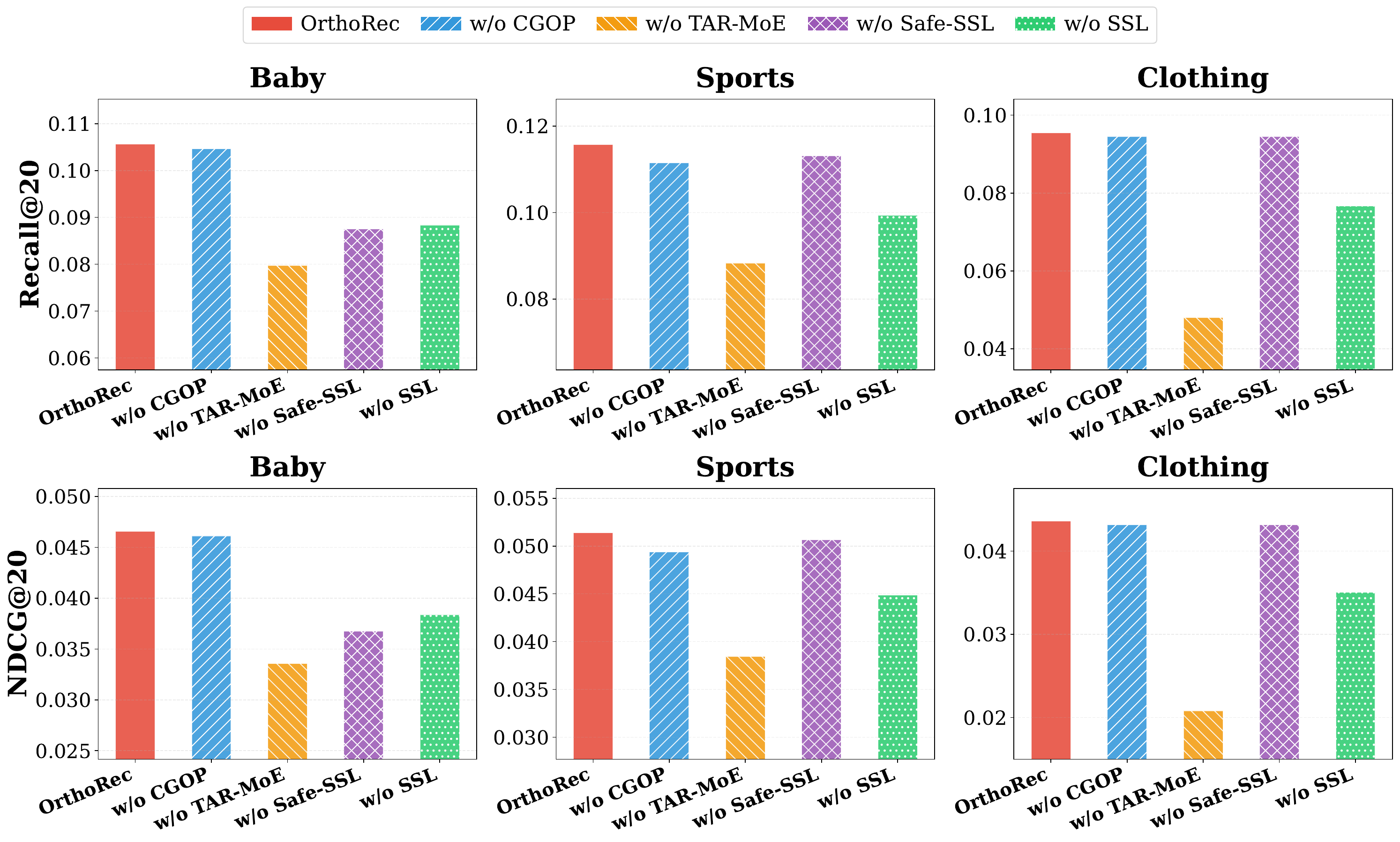} 
    \caption{Ablation study on OrthoRec's key design choices.}
    \label{fig:ablation}
\end{figure}

To validate our design choices, we compare OrthoRec against four variants: (1) \textbf{w/o CGOP Truncation} disables the adaptive truncation gate ($\gamma=1$), (2) \textbf{w/o TAR-MoE (softmax)} replaces the decoupled Sigmoid routing with standard softmax, (3) \textbf{w/o safe-SSL} degrades our conflict-aware penalty to standard InfoNCE, and (4) \textbf{w/o SSL} removes the self-supervised objective entirely. Based on Figure~\ref{fig:ablation}, we conclude:

\textbf{(1) Decoupled routing avoids zero-sum competition bottlenecks.} Replacing TAR-MoE with softmax is the single most damaging ablation on every dataset, and especially severe on \textit{Clothing}. The effect stems from the constraint itself rather than injection magnitude: re-tuning $\gamma$ for the softmax variant on a magnitude-matched grid selects the same scale OrthoRec already uses. Forcing heterogeneous modalities into mutually exclusive distributions thus limits joint expressiveness, whereas independent routing is crucial for robust fusion.
\textbf{(2) Adaptive truncation and decoupled routing are complementary.} Routing controls how much purified signal enters, truncation controls which directions. Disabling truncation isolates the directional effect alone, which is modest on curated benchmark features. This ablation is thus a lower bound on CGOP's contribution: uncorrupted features carry little deceptive component to remove, and directional filtering matters once conflicts intensify, as Section~\ref{sec:robustness} confirms.
\textbf{(3) Cross-modal alignment is indispensable and must be conflict-aware.} Removing SSL entirely (\textit{w/o SSL}) drops R@20 by 14.1--19.7\% across all datasets, so alignment carries a large share of the gain. Degrading it to standard InfoNCE (\textit{w/o safe-SSL}) separates how alignment is applied from whether it is: on \textit{Baby} the unguarded objective recovers none of the gap, so forcing contradictory pairs together is as damaging as no alignment at all. Elsewhere it operates near its safety margin at the tuned weight, with Section~\ref{sec:rq5} showing sharp degradation once the weight exceeds that margin.

\subsection{Robustness against Modality Noise (RQ3)}
\label{sec:robustness}

\begin{figure}[t!] 
    \centering
    \includegraphics[width=\columnwidth]{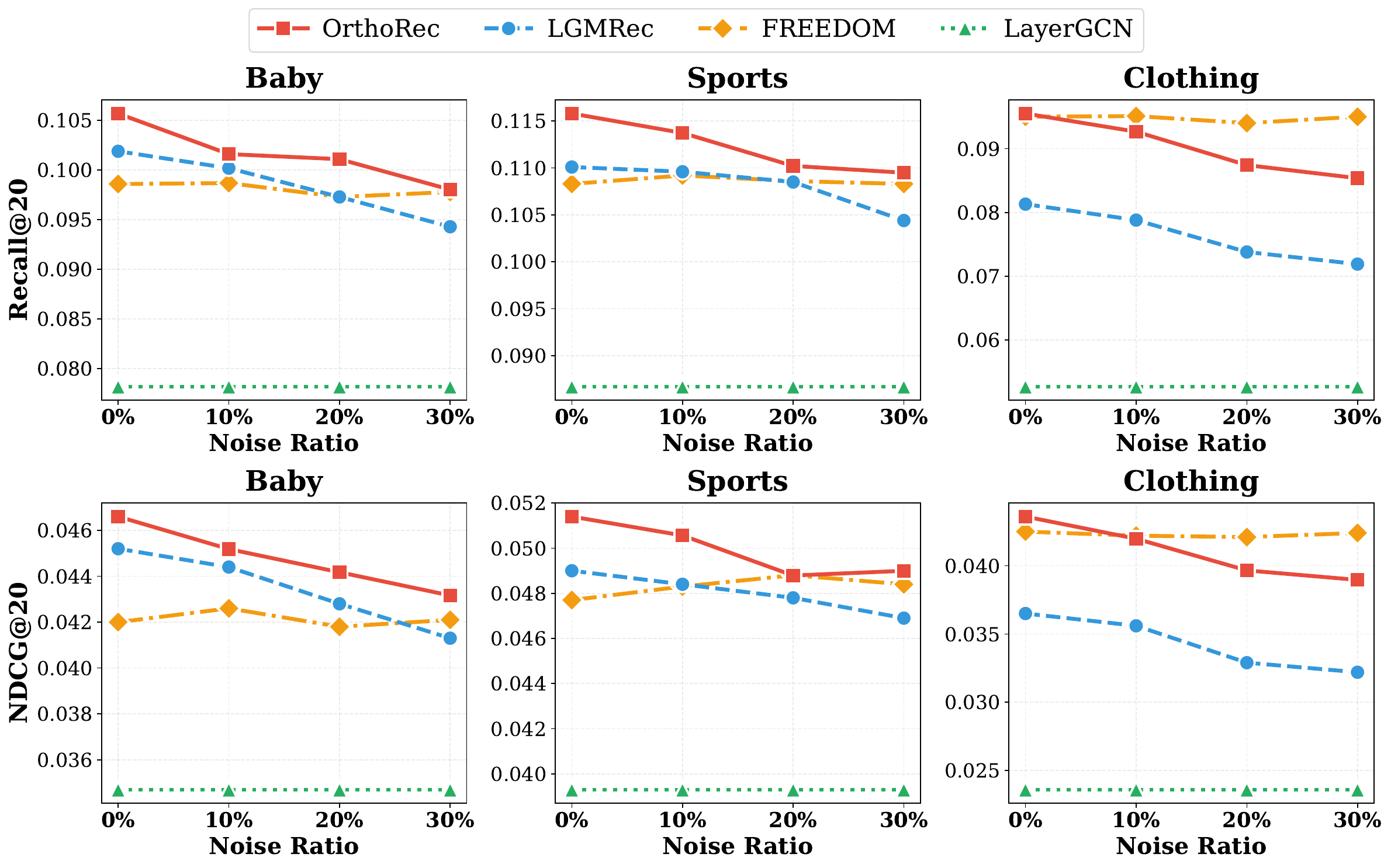} 
    \caption{Performance evaluation under varying visual noise ratios.}
    \label{fig:noise_curve}
\end{figure}

To evaluate robustness against modality-topology conflicts, we inject synthetic noise by randomly shuffling visual features for a proportion ($p \in \{0\%, 10\%, 20\%, 30\%\}$) of items during both training and inference. Figure~\ref{fig:noise_curve} compares OrthoRec against LayerGCN (pure CF), FREEDOM (graph-based), and LGMRec (deep-fusion). The trajectories expose a utilization-robustness trade-off that OrthoRec substantially alleviates:
\textbf{(1) A flat curve measures modality reliance, not robustness.}
LayerGCN is perfectly horizontal because it uses no visual features, yet trails OrthoRec by $0.020$--$0.033$ Recall@20 at every ratio. FREEDOM is likewise near-flat ($\leq 0.8\%$ degradation) as its frozen item-item graph weights the visual channel by only $0.1$. Flatness here comes from declining to use the modality, and reading it as noise tolerance would reward the very conservatism our method avoids.
\textbf{(2) Unfiltered deep fusion converts modality signal into a liability.}
LGMRec integrates modalities aggressively and is competitive on clean data, but degrades on every dataset and stays below OrthoRec at every ratio, by a margin that holds essentially constant rather than shrinking. As the two models see identical features and differ mainly in whether the signal is purified and decoupled, na\"ive early fusion evidently amplifies toxic noise instead of filtering it.
\textbf{(3) OrthoRec exploits semantics without inheriting their fragility.}
OrthoRec attains the highest clean-data accuracy on all three datasets and, against the comparable deep-fusion baseline, both starts higher and falls more slowly ($10.6\%$ relative R@20 drop at $30\%$ noise versus $11.6\%$). Its residual degradation is the expected cost of genuinely using the visual channel; what CGOP buys is a bound on corruption propagation, since truncating the anchor-orthogonal component removes the directions along which shuffled features would otherwise pollute the collaborative space. The trade-off curve thus shifts rather than tilts. As this protocol shuffles visual features, robustness to naturally occurring corruption is left for future work.

\subsection{Item Popularity and Sparsity Analysis (RQ4)}
\label{sec:sparsity}

\begin{figure}[t!] 
    \centering
    \includegraphics[width=\columnwidth]{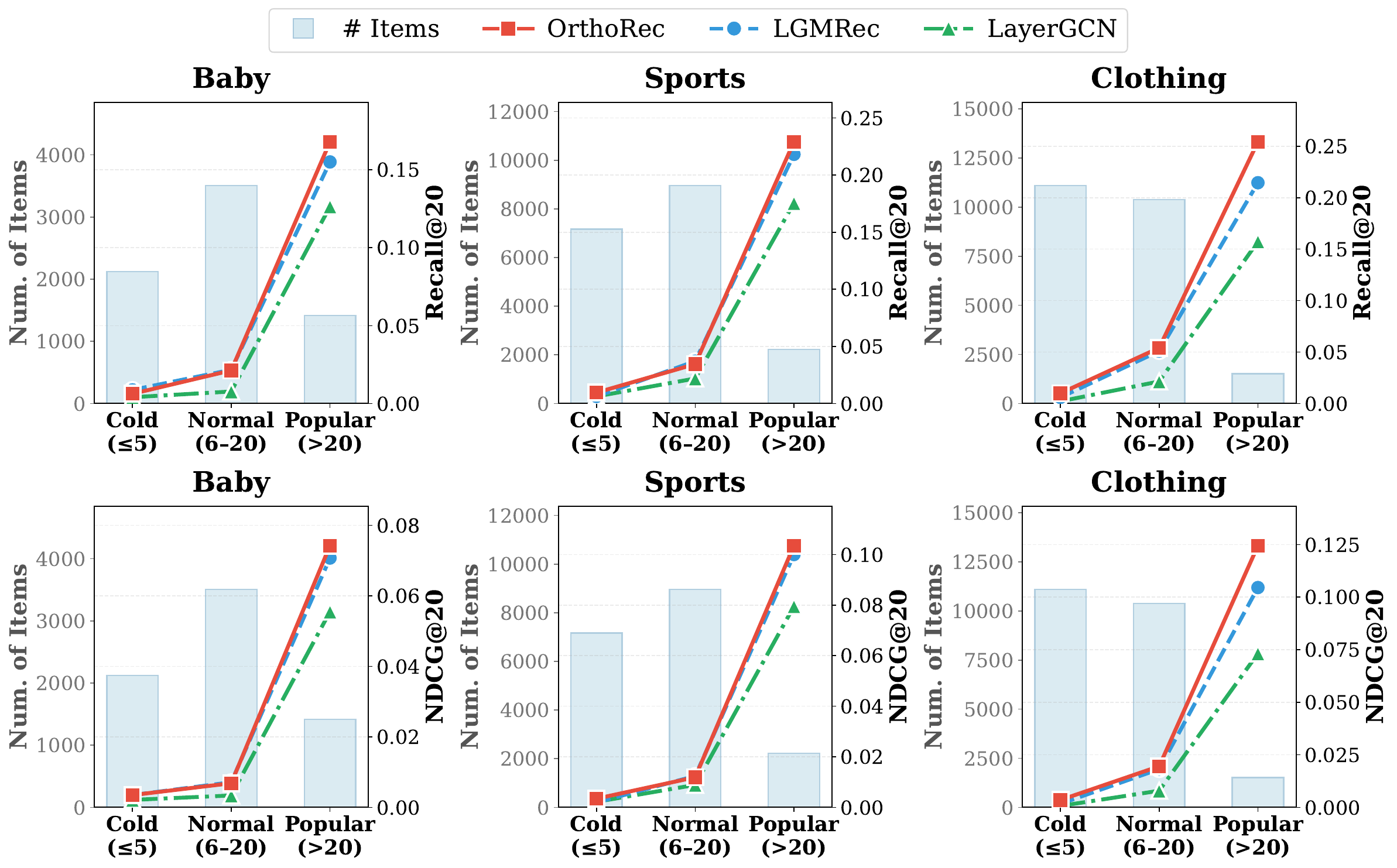} 
    \caption{Recommendation performance across item popularity groups.} 
    \label{fig:sparsity}
\end{figure}

To evaluate sparsity handling, we partition test items by training frequency: \textit{Cold} ($\leq 5$), \textit{Normal} ($6 \sim 20$), and \textit{Popular} ($> 20$). Figure~\ref{fig:sparsity} compares OrthoRec against LGMRec and LayerGCN, revealing key insights:
\textbf{(1) Complementary strengths across popularity groups.}
OrthoRec's most pronounced advantage lies in the \textit{Cold} group (about +41\% R@20 over LGMRec), while the two models stay within $4\%$ on \textit{Popular} items, where abundant collaborative edges already yield high-quality CF representations. The gain is thus concentrated where collaborative evidence is scarce rather than spread uniformly across popularity levels.
\textbf{(2) Safe and robust utilization for cold items.}
For \textit{Cold} items with minimal structural edges, unfiltered early fusion is most vulnerable, as deceptive signals cannot be counterbalanced by reliable collaborative evidence, which is exactly where purification is most valuable. This matches the routing behaviour: since TAR-MoE conditions the injection scale on topological consensus, an item whose evidence is too thin to corroborate its features receives a conservative injection rather than an unverified one. On \textit{Baby}, the gains on \textit{Normal} and \textit{Popular} items show that this conservatism is confined to the regime where the anchor cannot yet certify the signal.
\textbf{(3) The indispensable role of multimodal semantics.}
The pure CF model (LayerGCN) consistently underperforms, nearly collapsing on \textit{Cold} items, confirming that purified multimodal semantics remain essential to overcome structural sparsity.

\begin{figure}[t]
\centering
\includegraphics[width=\columnwidth]{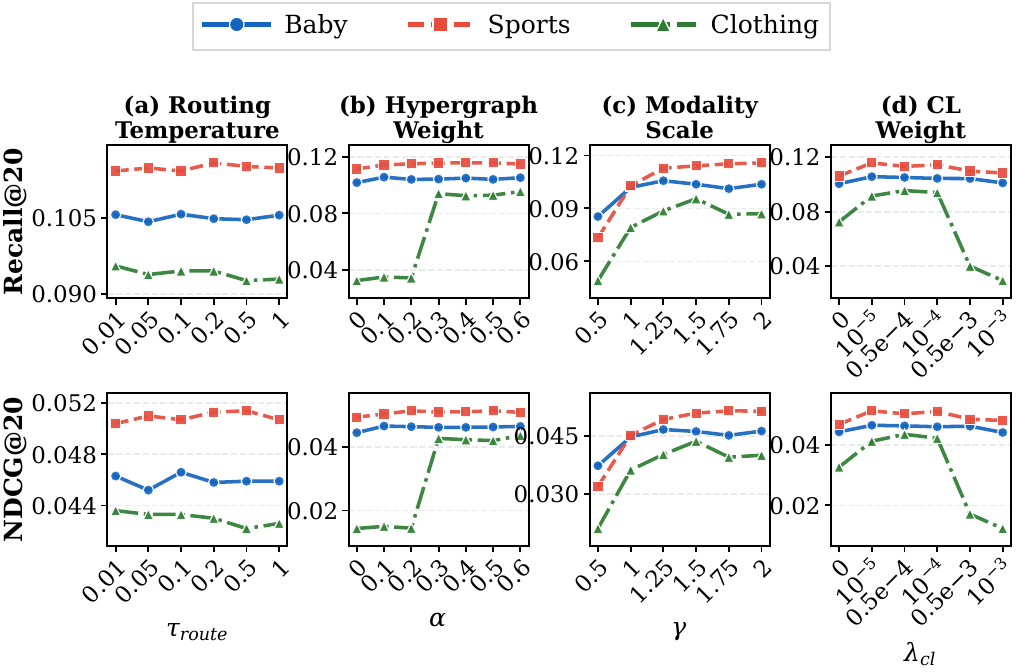}
\caption{OrthoRec's hyperparameter sensitivity on three datasets.}
\label{fig:sensitivity}
\vspace{-0.1cm}
\end{figure}


\subsection{Hyperparameter Sensitivity (RQ5)}
\label{sec:rq5}

To evaluate OrthoRec's robustness, we investigate its sensitivity to four key hyperparameters in Figure~\ref{fig:sensitivity}:

\textbf{Impact of $\tau_{r}$.}
OrthoRec is stable across $\tau_{r} \in [0.01, 1.0]$, with fluctuation below $4\%$. Since $\tau_r$ only sharpens the sigmoid on the routing logits, the gates keep comparable relative preferences over a wide range, so no delicate tuning is required.

\textbf{Impact of $\alpha$.}
$\alpha$ controls the global hypergraph enhancement injected into the collaborative anchor. \textit{Baby} and \textit{Sports} are stable across the range, while \textit{Clothing} requires $\alpha \geq 0.3$. This follows from the anchor's role: \textit{Clothing} is the sparsest dataset, so local edges alone cannot yield a trustworthy geometric reference and the global hypergraph must supply the missing structure before purification can decide which directions to keep. The dependence thus confirms the mechanism rather than revealing fragility, and $\alpha = 0.5$ sits safely inside the stable region on all three datasets.

\textbf{Impact of $\gamma$.}
Performance improves sharply once $\gamma$ exceeds $1$, then flattens, peaking within $\gamma \in[1.25, 1.5]$. Multimodal semantics thus matter, yet excessive amplification introduces noise that overpowers collaborative signals.

\textbf{Impact of $\lambda_{cl}$.}
A moderate contrastive signal ($\lambda_{cl} \in [10^{-5}, 10^{-4}]$) yields the best alignment, and removing it ($\lambda_{cl}=0$) is clearly worse everywhere, matching Section~\ref{sec:ablation}. Pushing past this range ($\lambda_{cl} \ge 5 \times 10^{-4}$) degrades \textit{Clothing} markedly, the ``Latent Distortion Dilemma'' the conflict score is designed to contain. The best range is broad and consistent across datasets, so the sensitivity reflects alignment pressure rather than a brittle setting.

\subsection{Qualitative Analysis (RQ6)}
\label{sec:rq6}

\begin{figure}[t]
\centering
\includegraphics[width=\columnwidth]{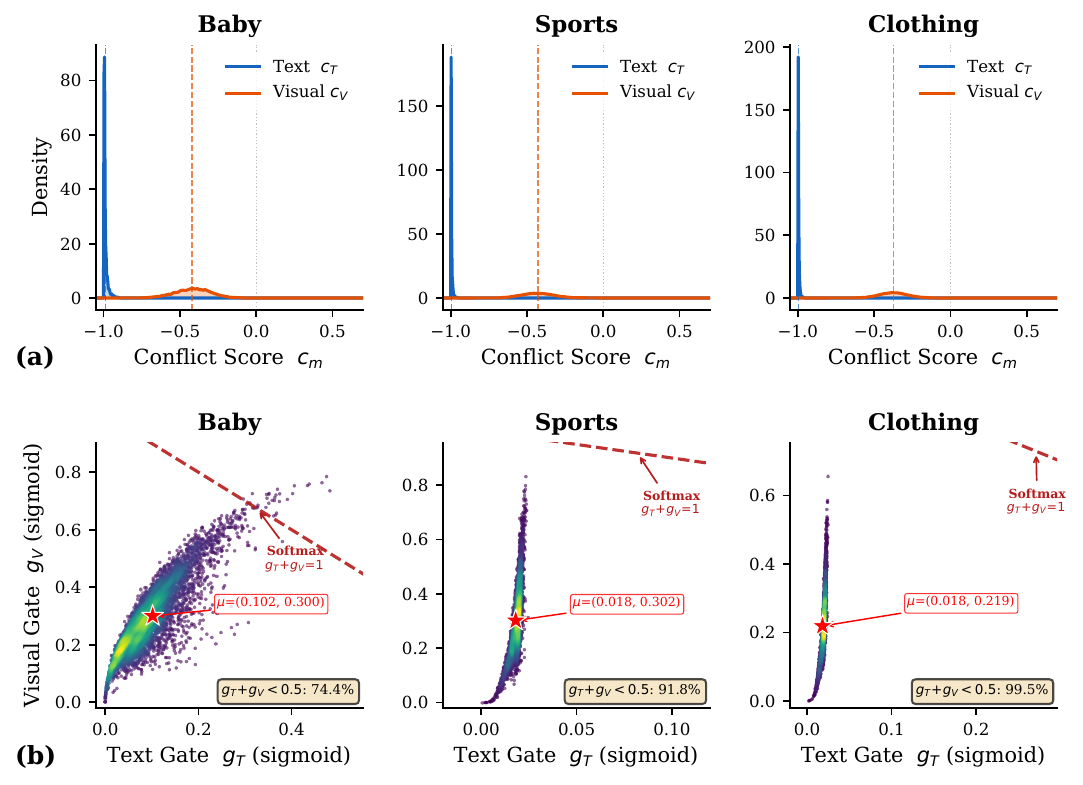}
\caption{Visualizations of (a) latent conflict scores and (b) TAR-MoE routing weights across the three datasets.}
\label{fig:distribution}
\end{figure}

To understand OrthoRec's inner workings, we visualize the latent distributions and gating weights in Figure~\ref{fig:distribution}.
\textbf{(1) Heterogeneous modality-topology conflicts.}
Raw conflict scores $c_m = \cos(\mathbf{e}_m, \mathbf{e}_{CF})$ (top row) reveal severe geometric divergence between pre-trained semantics and the CF space. Textual scores concentrate near $-0.99$, while visual scores spread from below $-0.8$ up to $+0.56$ on \textit{Clothing}, confirming that raw modalities suffer varying spatial misalignments and rendering static fusion suboptimal.
\textbf{(2) Intelligent suppression via decoupled gating.}
The bottom row plots TAR-MoE routing weights $(g_T, g_V)$. Softmax MoE enforces $g_T + g_V = 1$, so a toxic modality can only be suppressed by promoting another rather than being filtered on its own merits, consistent with the degradation in Figure~\ref{fig:ablation}. Our decoupled sigmoid gating instead acts as an independent filter: facing extreme textual conflict it suppresses text ($g_T \sim 0.02$) while visual weights span up to $0.83$. The learned gate sums stay well below the softmax diagonal (averaging $0.24$--$0.40$ across datasets), with ${>}99\%$ of \textit{Clothing} items satisfying $g_T + g_V < 0.5$. Breaking the zero-sum constraint thus grants the freedom to suppress multiple noisy modalities at once.

\begin{figure}[t]
    \centering
    \includegraphics[width=\columnwidth]{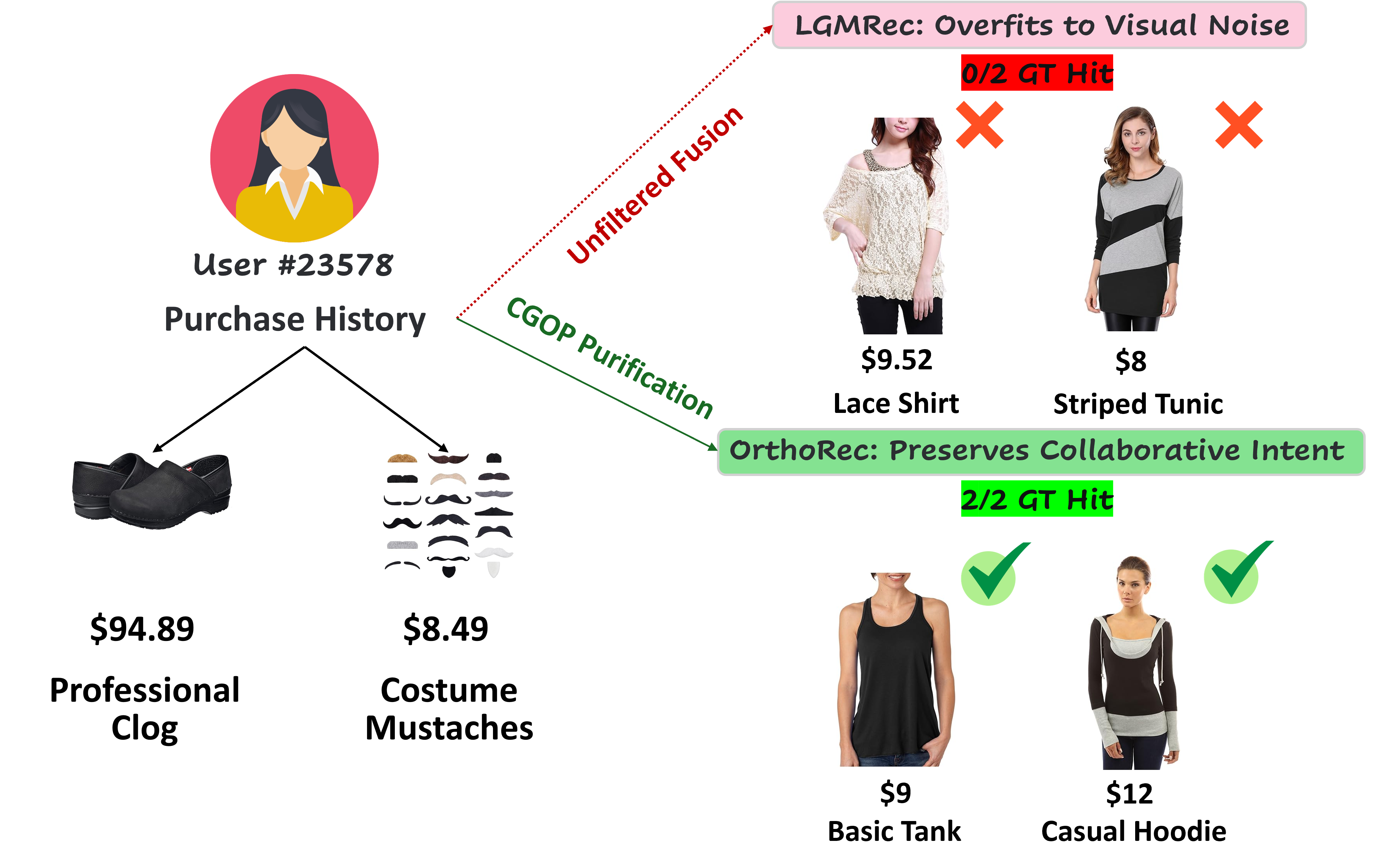}
    \caption{Case study of User \#23578.}
    \label{fig:case_study}
\end{figure}

\textbf{Case Study.}
Figure~\ref{fig:case_study} compares both models on \textit{Clothing} for User~\#23578. Although \textit{Clothing} shows the highest visual scores overall, this user sits in the negative tail: their images conflict sharply with the collaborative topology and act as deceptive noise. Such per-item variation is what item-wise truncation and topology-conditioned routing exist to handle, since a dataset-level weight would inherit the average rather than this user's conflict. LGMRec's unfiltered fusion overfits to appealing but irrelevant visual semantics and misses both targets, whereas CGOP truncates the toxic components, shielding the user's true intent and retrieving both targets in the top-20.

\section{Conclusion}
We propose OrthoRec to alleviate the modality-topology conflicts that distort latent spaces in multimodal recommendation. By unifying Collaborative-Guided Orthogonal Purification (CGOP) for geometric noise filtration, decoupled TAR-MoE routing, and safe-SSL, OrthoRec delivers consistent accuracy improvements over competitive baselines and improved robustness under modality noise and item sparsity within the evaluated settings.

\begin{acks}
This work was supported by the Strategic Support Fund of City University of Hong Kong under Project 7020230, and by the National Natural Science Foundation of China under Grant 62302420.
\end{acks}

\bibliographystyle{ACM-Reference-Format}
\bibliography{sample-base}

@inproceedings{he2016vbpr,
  title={VBPR: visual bayesian personalized ranking from implicit feedback},
  author={He, Ruining and McAuley, Julian},
  booktitle={Proceedings of the AAAI conference on artificial intelligence},
  volume={30},
  number={1},
  year={2016}
}

@inproceedings{mcauley2015image,
  title={Image-based recommendations on styles and substitutes},
  author={McAuley, Julian and Targett, Christopher and Shi, Qinfeng and Van Den Hengel, Anton},
  booktitle={Proceedings of the 38th international ACM SIGIR conference on research and development in information retrieval},
  pages={43--52},
  year={2015}
}

@article{rendle2012bpr,
  title={BPR: Bayesian personalized ranking from implicit feedback},
  author={Rendle, Steffen and Freudenthaler, Christoph and Gantner, Zeno and Schmidt-Thieme, Lars},
  journal={arXiv preprint arXiv:1205.2618},
  year={2012}
}

@inproceedings{zheng2021cold,
  title={Cold-start Sequential Recommendation via Meta Learner},
  author={Zheng, Yujia and Liu, Siyi and Li, Zekun and Wu, Shu},
  booktitle={Proceedings of the Thirty-Fifth AAAI Conference on Artificial Intelligence},
  pages={4706--4713},
  year={2021}
}

@inproceedings{wei2019mmgcn,
  title={MMGCN: Multi-modal Graph Convolution Network for Personalized Recommendation of Micro-video},
  author={Wei, Yinwei and Wang, Xiang and Nie, Liqiang and He, Xiangnan and Hong, Richang and Chua, Tat-Seng},
  booktitle={Proceedings of the 27th ACM International Conference on Multimedia},
  pages={1437--1445},
  year={2019}
}

@article{tao2022self,
  title={Self-supervised learning for multimedia recommendation},
  author={Tao, Zhulin and Liu, Xiaohao and Xia, Yewei and Wang, Xiang and Yang, Lifang and Huang, Xianglin and Chua, Tat-Seng},
  journal={IEEE Transactions on Multimedia},
  volume={25},
  pages={5107--5116},
  year={2022},
  publisher={IEEE}
}

@inproceedings{zhou2023bootstrap,
  title={Bootstrap Latent Representations for Multi-modal Recommendation},
  author={Zhou, Xin and Zhou, Hongyu and Liu, Yong and Zeng, Zhiwei and Miao, Chunyan and Wang, Pengwei and You, Yuan and Jiang, Feijun},
  booktitle={Proceedings of the ACM Web Conference 2023},
  pages={845--854},
  year={2023}
}

@inproceedings{xu2025mentor,
  title={Mentor: multi-level self-supervised learning for multimodal recommendation},
  author={Xu, Jinfeng and Chen, Zheyu and Yang, Shuo and Li, Jinze and Wang, Hewei and Ngai, Edith CH},
  booktitle={Proceedings of the AAAI Conference on Artificial Intelligence},
  volume={39},
  number={12},
  pages={12908--12917},
  year={2025}
}

@inproceedings{he2020lightgcn,
  title={Lightgcn: Simplifying and powering graph convolution network for recommendation},
  author={He, Xiangnan and Deng, Kuan and Wang, Xiang and Li, Yan and Zhang, Yongdong and Wang, Meng},
  booktitle={Proceedings of the 43rd International ACM SIGIR conference on research and development in Information Retrieval},
  pages={639--648},
  year={2020}
}

@article{liu2024multimodal,
  title={Multimodal recommender systems: A survey},
  author={Liu, Qidong and Hu, Jiaxi and Xiao, Yutian and Zhao, Xiangyu and Gao, Jingtong and Wang, Wanyu and Li, Qing and Tang, Jiliang},
  journal={ACM Computing Surveys},
  volume={57},
  number={2},
  pages={1--17},
  year={2024}
}

@article{zhang2022sedgn,
  title={SEDGN: Sequence enhanced denoising graph neural network for session-based recommendation},
  author={Zhang, Chunkai and Zheng, Wenjing and Liu, Quan and Nie, Junli and Zhang, Hanyu},
  journal={Expert Systems with Applications},
  volume={203},
  pages={117391},
  year={2022}
}

@article{guo2023noise,
  title={Pace-Adaptive and Noise-Resistant Contrastive Learning for Multimodal Feature Fusion},
  author={Guo, Xiaobao and Kot, Alex and Kong, Adams Wai-Kin},
  journal={IEEE Transactions on Multimedia},
  volume={25},
  pages={9437-9448},
  year={2023}
}

@inproceedings{jiang2024diffmm,
  title={Diffmm: Multi-modal diffusion model for recommendation},
  author={Jiang, Yangqin and Xia, Lianghao and Wei, Wei and Luo, Da and Lin, Kangyi and Huang, Chao},
  booktitle={Proceedings of the 32nd ACM international conference on multimedia},
  pages={7591--7599},
  year={2024}
}

@article{gao2024causal,
  title={Causal inference in recommender systems: A survey and future directions},
  author={Gao, Chen and Zheng, Yu and Wang, Wenjie and Feng, Fuli and He, Xiangnan and Li, Yong},
  journal={ACM Transactions on Information Systems},
  volume={42},
  number={4},
  pages={1--32},
  year={2024}
}

@inproceedings{mu2022alleviating,
  title={Alleviating spurious correlations in knowledge-aware recommendations through counterfactual generator},
  author={Mu, Shanlei and Li, Yaliang and Zhao, Wayne Xin and Wang, Jingyuan and Ding, Bolin and Wen, Ji-Rong},
  booktitle={Proceedings of the 45th international ACM SIGIR conference on research and development in information retrieval},
  pages={1401--1411},
  year={2022}
}

@inproceedings{chen2020simple,
  title={A simple framework for contrastive learning of visual representations},
  author={Chen, Ting and Kornblith, Simon and Norouzi, Mohammad and Hinton, Geoffrey},
  booktitle={International conference on machine learning},
  pages={1597--1607},
  year={2020}
}

@inproceedings{yu2022graph,
  title={Are graph augmentations necessary? simple graph contrastive learning for recommendation},
  author={Yu, Junliang and Yin, Hongzhi and Xia, Xin and Chen, Tong and Cui, Lizhen and Nguyen, Quoc Viet Hung},
  booktitle={Proceedings of the 45th international ACM SIGIR conference on research and development in information retrieval},
  pages={1294--1303},
  year={2022}
}

@inproceedings{wang2020understanding,
  title={Understanding contrastive representation learning through alignment and uniformity on the hypersphere},
  author={Wang, Tongzhou and Isola, Phillip},
  booktitle={International conference on machine learning},
  pages={9929--9939},
  year={2020}
}

@article{robinson2020contrastive,
  title={Contrastive learning with hard negative samples},
  author={Robinson, Joshua and Chuang, Ching-Yao and Sra, Suvrit and Jegelka, Stefanie},
  journal={arXiv preprint arXiv:2010.04592},
  year={2020}
}

@inproceedings{yang2023debiased,
  title={Debiased contrastive learning for sequential recommendation},
  author={Yang, Yuhao and Huang, Chao and Xia, Lianghao and Huang, Chunzhen and Luo, Da and Lin, Kangyi},
  booktitle={Proceedings of the ACM web conference 2023},
  pages={1063--1073},
  year={2023}
}

@article{tian2020makes,
  title={What makes for good views for contrastive learning?},
  author={Tian, Yonglong and Sun, Chen and Poole, Ben and Krishnan, Dilip and Schmid, Cordelia and Isola, Phillip},
  journal={Advances in neural information processing systems},
  volume={33},
  pages={6827--6839},
  year={2020}
}

@article{vaswani2017attention,
  title={Attention is all you need},
  author={Vaswani, Ashish and Shazeer, Noam and Parmar, Niki and Uszkoreit, Jakob and Jones, Llion and Gomez, Aidan N and Kaiser, {\L}ukasz and Polosukhin, Illia},
  journal={Advances in neural information processing systems},
  volume={30},
  year={2017}
}

@inproceedings{peng2022balanced,
  title={Balanced multimodal learning via on-the-fly gradient modulation},
  author={Peng, Xiaokang and Wei, Yake and Deng, Andong and Wang, Dong and Hu, Di},
  booktitle={Proceedings of the IEEE/CVF conference on computer vision and pattern recognition},
  pages={8238--8247},
  year={2022}
}

@article{shazeer2017outrageously,
  title={Outrageously large neural networks: The sparsely-gated mixture-of-experts layer},
  author={Shazeer, Noam and Mirhoseini, Azalia and Maziarz, Krzysztof and Davis, Andy and Le, Quoc and Hinton, Geoffrey and Dean, Jeff},
  journal={arXiv preprint arXiv:1701.06538},
  year={2017}
}

@inproceedings{ma2018modeling,
  title={Modeling task relationships in multi-task learning with multi-gate mixture-of-experts},
  author={Ma, Jiaqi and Zhao, Zhe and Yi, Xinyang and Chen, Jilin and Hong, Lichan and Chi, Ed H},
  booktitle={Proceedings of the 24th ACM SIGKDD international conference on knowledge discovery \& data mining},
  pages={1930--1939},
  year={2018}
}

@inproceedings{wei2020grcn,
  title={Graph-Refined Convolutional Network for Multimedia Recommendation with Implicit Feedback},
  author={Wei, Yinwei and Wang, Xiang employment and Nie, Liqiang and He, Xiangnan and Chua, Tat-Seng},
  booktitle={Proceedings of the 28th ACM International Conference on Multimedia},
  pages={3541--3549},
  year={2020}
}

@inproceedings{zhang2021mining,
  title={Mining Latent Structures for Multimedia Recommendation},
  author={Zhang, Jinghao and Zhu, Yanqiao and Liu, Qiang and Wu, Shu and Wang, Shuhui and Wang, Liang},
  booktitle={Proceedings of the 29th ACM International Conference on Multimedia},
  pages={3872--3880},
  year={2021}
}

@inproceedings{guo2024lgmrec,
  title={Lgmrec: Local and global graph learning for multimodal recommendation},
  author={Guo, Zhiqiang and Li, Jianjun and Li, Guohui and Wang, Chaoyang and Shi, Si and Ruan, Bin},
  booktitle={Proceedings of the AAAI conference on artificial intelligence},
  volume={38},
  number={8},
  pages={8454--8462},
  year={2024}
}

@inproceedings{zhou2023tale,
  title={A Tale of Two Graphs: Freezing and Denoising Graph Structures for Multimodal Recommendation},
  author={Zhou, Xin and Shen, Zhiqi},
  booktitle={Proceedings of the 31st ACM International Conference on Multimedia},
  pages={935--943},
  year={2023}
}

@inproceedings{zheng2021disentangling,
  title={Disentangling User Interest and Conformity for Recommendation with Causal Embedding},
  author={Zheng, Yu and Gao, Chen and Li, Xiang and He, Xiangnan and Li, Yong and Jin, Depeng},
  booktitle={Proceedings of the Web Conference 2021},
  pages={2980--2991},
  year={2021}
}

@article{wang2021dualgnn,
  title={Dualgnn: Dual graph neural network for multimedia recommendation},
  author={Wang, Qifan and Wei, Yinwei and Yin, Jianhua and Wu, Jianlong and Song, Xuemeng and Nie, Liqiang},
  journal={IEEE Transactions on Multimedia},
  volume={25},
  pages={1074--1084},
  year={2021}
}

@inproceedings{wang2024noise,
  title={Noise-resistant graph neural networks for session-based recommendation},
  author={Wang, Qi and Wu, Anbiao and Yuan, Ye and Wang, Yishu and Zhong, Guangqing and Gao, Xuefeng and Yang, Chenghu},
  booktitle={Asia-Pacific Web (APWeb) and Web-Age Information Management (WAIM) Joint International Conference on Web and Big Data},
  pages={144--160},
  year={2024}
}

@article{lin2025contrastive,
  title={Contrastive modality-disentangled learning for multimodal recommendation},
  author={Lin, Xixun and Liu, Rui and Cao, Yanan and Zou, Lixin and Li, Qian and Wu, Yongxuan and Liu, Yang and Yin, Dawei and Xu, Guandong},
  journal={ACM Transactions on Information Systems},
  volume={43},
  number={3},
  pages={1--31},
  year={2025}
}

@article{ma2019learning,
  title={Learning disentangled representations for recommendation},
  author={Ma, Jianxin and Zhou, Chang and Cui, Peng and Yang, Hongxia and Zhu, Wenwu},
  journal={Advances in neural information processing systems},
  volume={32},
  year={2019}
}

@article{rong2019dropedge,
  title={Dropedge: Towards deep graph convolutional networks on node classification},
  author={Rong, Yu and Huang, Wenbing and Xu, Tingyang and Huang, Junzhou},
  journal={arXiv preprint arXiv:1907.10903},
  year={2019}
}

@inproceedings{wang2023image,
  title={Image as a foreign language: Beit pretraining for vision and vision-language tasks},
  author={Wang, Wenhui and Bao, Hangbo and Dong, Li and Bjorck, Johan and Peng, Zhiliang and Liu, Qiang and Aggarwal, Kriti and Mohammed, Owais Khan and Singhal, Saksham and Som, Subhojit and others},
  booktitle={Proceedings of the IEEE/CVF Conference on Computer Vision and Pattern Recognition},
  pages={19175--19186},
  year={2023}
}

@inproceedings{zhou2023layer,
  title={Layer-refined graph convolutional networks for recommendation},
  author={Zhou, Xin and Lin, Donghai and Liu, Yong and Miao, Chunyan},
  booktitle={2023 IEEE 39th International Conference on Data Engineering (ICDE)},
  pages={1247--1259},
  year={2023}
}

@article{chen2024bge,
  title={Bge m3-embedding: Multi-lingual, multi-functionality, multi-granularity text embeddings through self-knowledge distillation},
  author={Chen, Jianlv and Xiao, Shitao and Zhang, Peitian and Luo, Kun and Lian, Defu and Liu, Zheng},
  journal={arXiv preprint arXiv:2402.03216},
  volume={4},
  number={5},
  year={2024}
}

@inproceedings{liu2026motorec,
  title={MoToRec: Sparse-Regularized Multimodal Tokenization for Cold-Start Recommender},
  author={Liu, Jialin and Zhang, Zhaorui and Cheung, Ray CC},
  booktitle={Proceedings of the AAAI Conference on Artificial Intelligence},
  volume={40},
  number={18},
  pages={15324--15332},
  year={2026}
}

@inproceedings{jiang2024mmgcl,
  title={Mmgcl: Meta knowledge-enhanced multi-view graph contrastive learning for recommendations},
  author={Jiang, Yuezihan and Li, Changyu and Chen, Gaode and Li, Peiyi and Zhang, Qi and Lin, Jingjian and Jiang, Peng and Sun, Fei and Zhang, Wentao},
  booktitle={Proceedings of the 18th ACM conference on recommender systems},
  pages={538--548},
  year={2024}
}

@inproceedings{xv2024damrs,
  title={Improving multi-modal recommender systems by denoising and aligning multi-modal content and user feedback},
  author={Xv, Guipeng and Li, Xinyu and Xie, Ruobing and Lin, Chen and Liu, Chong and Xia, Feng and Kang, Zhanhui and Lin, Leyu},
  booktitle={Proceedings of the 30th ACM SIGKDD conference on knowledge discovery and data mining},
  pages={3645--3656},
  year={2024}
}

@inproceedings{liu2026sga,
  title={SGA-GNN: Semantic-Guided Adaptive Graph Neural Network for Cold-Start Multimodal Recommendation},
  author={Liu, Jialin and Zhang, Zhaorui and Cheung, Ray CC},
  booktitle={ICASSP 2026-2026 IEEE International Conference on Acoustics, Speech and Signal Processing (ICASSP)},
  pages={1266--1270},
  year={2026},
  organization={IEEE}
}

\end{document}